\documentclass[aps,prb,nofootinbib]{revtex4}
\usepackage{graphicx}
\usepackage{amsfonts}
\usepackage{amsmath}

\newcommand{\be}{\begin{eqnarray}}
\newcommand{\ee}{\end{eqnarray}}

\newcommand{\x}{\tilde x}
\newcommand{\om}{\omega}
\newcommand{\Om}{\Omega}
\newcommand{\Sb}{(b\Om, \Om)}

\begin{document}

\author{O. Corradini}
\email{olindo.corradini@bo.infn.it}
\affiliation{Dipartimento di Fisica, Universit\`a degli Studi di Bologna and I.N.F.N. 
 Sezione di Bologna,
Via Irnerio 46, Bologna, I-40126  Italy}
\author{ P. Faccioli}
\email{faccioli@science.unitn.it}
\affiliation{Dipartimento di Fisica,  Universit\`a degli Studi di
  Trento and I.N.F.N, Via Sommarive 14, Povo (Trento), I-38050 Italy} 
\affiliation{Institut de Physique Th\'eorique,
Centre d'Etudes de Saclay, CEA, IPhT, F-91191, Gif-sur-Yvette, France}
\author{H. Orland}
\email{henri.orland@cea.fr}
\affiliation{Institut de Physique Th\'eorique,
Centre d'Etudes de Saclay, CEA, IPhT, F-91191, Gif-sur-Yvette, France}
\begin{abstract} 

We present a novel approach to  investigate the long-time stochastic
dynamics of multi-dimensional classical systems, in contact with a
heat-bath. When the potential energy landscape is rugged, the kinetics 
 displays a decoupling of short and long time scales and both
 Molecular Dynamics (MD)  or Monte Carlo (MC) simulations are generally  inefficient.
Using  a field theoretic approach, we perform analytically the average over the short-time stochastic fluctuations. This way, we obtain an effective theory, which  generates the same long-time dynamics of the original theory, 
but has a lower time resolution power. 
Such an approach  is used to develop an improved version of the MC algorithm, which is particularly suitable to investigate the dynamics of rare conformational transitions.
In the specific case of molecular systems at room temperature, we show that elementary integration time steps  used  to simulate the effective theory can be chosen a factor $\sim 100$ larger than those used  
in the original theory. Our results are illustrated and tested on a
simple system, characterized by a rugged energy landscape.   
\end{abstract}
\title{Simulating Stochastic Dynamics Using Large Time Steps}
\maketitle

\section{Introduction}

The investigation of a vast class of physical phenomena 
requires the  understanding of the long-time dynamics of classical systems, in contact with a heat-bath. Examples include critical dynamics, molecular aggregation, protein folding, to name a few.

The most natural strategy to describe these processes is to integrate numerically the equations of motion, i.e. to perform  MD simulations. 
Unfortunately, when the number of degrees of freedom  is very large, or in the presence of large free energy barriers,  MD approaches become extremely costly~\cite{pandeeCo}, or even impracticable. 
The problem arises  because the time scale associated 
with the system's local conformational changes can be many orders of magnitude smaller that the time scales of the dynamics one is interesting in studying. As a result, most of the computational time is invested in simulating uninteresting thermal oscillations. 

\begin{figure}[t]
\includegraphics[width=7cm, height=8.5 cm]{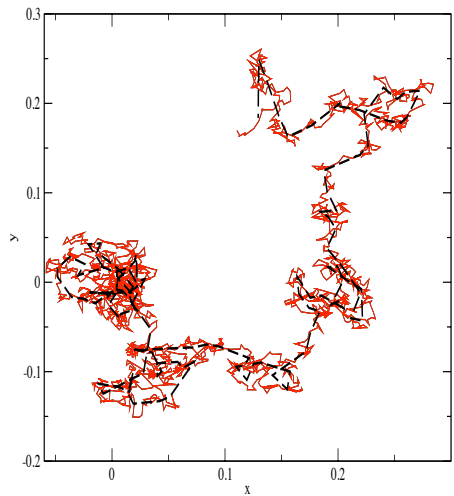}
\caption{Langevin dynamics of a point particle in a 2-dimensional external potential. The solid line denotes the result of an MD simulation. The dashed line is the result of averaging over blocks of consecutive frames of the MD trajectory.
Such an average smoothes out the trajectory.}
\label{exFig}
\end{figure}

This situation is exemplified in Fig.\ref{exFig}, where we show the stochastic motion of a point particle, 
interacting with a 2-dimensional external potential. The solid line was obtained by  means of a MD simulation and illustrates how,  at short time scales,  the dynamics of this system is dominated by fast modes associated
to  thermal diffusion. 
However,  when the evolution of the system is described using much lower time resolution power, the effect of such short-time thermal fluctuations tends to average out and to become unimportant. 
This is evident from the comparison between the solid line and the dashed line, which was obtained by averaging over blocks of consecutive frames in the original MD trajectory.  At long times, the dynamics of system is mostly sensitive to 
the  structure of the external energy landscape, which was chosen to be spherically symmetric.

Clearly, an important question to ask is whether it is possible to develop  theoretical/computational frameworks which yield directly the correct long-dynamics, but avoid investing computational time in 
simulating the short-time thermal oscillations. Significant progress in this direction has been recently made developing  approaches based on Markov State Models~\cite{MSM1,MSM2,MSM3}. 
A potential difficulty of such approaches  resides in correct identification of the metastable states. In addition, for each different system,  one needs  to perform a large set of independent MD simulations
in order to accurate calculation of the rate coefficients.

In this work, we present an alternative approach to simulate  the  dynamics over long times. We develop a rigorous effective theory which (i) yields by construction the correct long-dynamics and (ii) does not require  to identify meta-stable states, nor to evaluate
the transition matrix by MD.  To our goal,  we use a field theory approach, based on Renormalization Group (RG) ideas and on the notion of effective field theory~\cite{howto}.  
Such a powerful tools have been already successfully applied to describe the low-energy 
dynamics of a  vast variety of    of quantum and statistical systems
characterized by a separation of scales~---see e.g. \cite{zinn, kleinert}---. To
the best of our knowledge, this method has never been  applied to develop an effective theory to efficiently simulate
the long-time stochastic  dynamics of a system in contact with a heat
bath. 
  
The main idea of our  approach is to exploit the decoupling of time scales in the system in order  to define a perturbative series, in which  the expansion parameter  is the ratio of  short- over large- time scales. 
In such a perturbative framework, the average over the short-time fluctuations can be computed  analytically, to any desired level of accuracy. 
The average over the fast thermal oscillations gives rise to new terms in the stochastic path integral, which represent corrections  both to the interaction and to the diffusion coefficient. 
Such new terms implicitly take into account of the dynamics of the fast degrees of freedom, which have been integrated out from the system.  

Once a finite number of  such effective terms corresponding to a given accuracy have been calculated analytically, it is possible to simulate the dynamics of the system using much larger time steps.
By construction,  in  the regime of decoupling of fast and slow modes,  one is guaranteed that  the effective long time theory generates the same probability distributions of the underlying,  more fundamental  stochastic theory. 
It is important to emphasize the fact that the present approach is not equivalent to simply including higher-order corrections in the Trotter expansion~\cite{trotter}.
 Indeed, the assumption of decoupling of time scales leads to further simplifications with respect to such an approach.

The paper is organized as follows. In section \ref{sectionlang}, we review the path integral formulation of the  Langevin dynamics and we outline the formal connection between stochastic dynamics 
and evolution of a quantum particle in imaginary  time. Such a connection is used in section \ref{separation}  to identify and isolate the dynamics of the fast degrees of freedom.
In sections \ref{renormalized}  and \ref{perturbative} we present our perturbative scheme which allows to integrate out the fast modes and derive the effective interactions and diffusion coefficients. 
In section \ref{MC} we discuss how the effective theory for the dynamics at long time scales can be simulated using the diffusion MC algorithm, which is briefly reviewed in appendix \ref{algo}. Section \ref{example} is devoted to  simple  examples, which 
illustrate how this method works in practice. In section \ref{molecular} we discuss the applicability of the present approach to simulate the Langevin dynamics of molecular systems. Results and conclusions are summarized in section \ref{conclusions}. 

\section{Langevin Dynamics}
\label{sectionlang}
We consider a system defined by a stochastic $d$-dimensional variable $x$ obeying the Langevin Eq.n:
\be
\label{Lang}
m \ddot x = -\nabla U(x) - \gamma \dot x + \xi(t), 
\ee
where $U(x)$ is a potential energy function, $m$ is the mass, $\gamma$ is the friction coefficient and $\xi(t)$ is a $\delta-$correlated Gaussian noise.
In many molecular systems of interest, the acceleration term $m \ddot x$ is damped at time scales of the order $10^{-13}~s$, which much smaller than the time scale associated to local
conformational changes. If such a term is dropped one obtains the so-called \emph{over-damped} or  \emph{velocity Langevin Eq}.:
\be
\label{langevin}
\dot x = -\frac{1}{\gamma} \nabla U(x) + \eta(t),
\ee  
where  $\eta(t)$ is a rescaled delta-correlated Gaussian noise, satisfying the fluctuation-dissipation relationship:
\be
\langle \eta(t') \eta(t) \rangle = 2 d \frac{1}{\beta~\gamma}~ \delta(t-t').
\ee

The over-damped Langevin Eq. defines a Markovian process. The probability distribution $P(x,t)$ generated by such a stochastic differential equation obeys the Fokker-Planck Eq.
\be
\label{FP}
\frac{\partial}{\partial t} P(x,t) = \frac{1}{\beta~\gamma}~ \nabla \left[ \nabla P(x,t) + \beta \nabla U(x) P (x,t)\right].
\ee
By performing the substitution 
\be
P(x,t)= e^{-\frac{\beta}{2 } U(x)}~\psi(x,t)
\label{sub}
\ee
the Fokker-Planck Eq.(\ref{FP}) can be recast in the form of a Schr\"odinger Equation in imaginary time:
\be
 -\frac{\partial}{\partial t} \psi(x,t) = \hat{H}_{eff}~\psi(x,t),
\label{SE}
\ee
where the effective "Quantum Hamiltonian" operator reads
\be
\hat{H}_{eff}~=~- \frac{1}{\beta \gamma}~\nabla^2 + \beta~ V_{eff}(x),
\label{Heff}
\ee
and  $V_{eff}(x)$ is called the effective potential and reads
\be
\label{Veff}
V_{eff}(x)= \frac{1}{4\gamma}\left( (\nabla U(x))^2- \frac 2 \beta \nabla^2 U(x)\right).
\ee
Hence, the problem of studying the diffusion of a classical particle  can be
 mapped into the problem of determining the quantum-mechanical propagation in imaginary time of a virtual system, defined by the effective quantum Hamiltonian (\ref{Heff}), interacting with  the  effective potential $V_{eff}(x)$.

Let $G(x_f, t_f| x_i)$ be the Green's function of the Fokker-Planck operator, subject to initial condition $x(0)=x_i$, i.e.
\be
\frac{\partial}{\partial t} G(x_f, t| x_i) - \frac{1}{\beta~\gamma}~ \nabla \left[ \nabla G(x_f, t| x_i) + \beta \nabla U G(x_f, t| x_i)\right] = \delta(t) \delta(x-x_i)
\ee
The  interpretation of such a Green's function is the probability for the system to be in $x$ at $t$, conditioned to start from   $x_i$ at the initial time.
Formally, such a conditional probability can be related to the "quantum" propagator of the effective Hamiltonian~(\ref{Heff}):
\be
G(x,t|x_i) &=&  \exp\left[-\frac{\beta}{2}(U(x)-U(x_i))\right]  ~K(x,t|x_i), \\
\label{G}
K(x,t|x_i) &=&   \langle x | e^{- t H_{eff} }| x_i\rangle.
\label{K}
\ee
Hence, it is immediate to derive a path integral representation of the Green's function $G(x,t|x_i)$:
\be
\label{GPI}
G(x,t|x_i) &=& e^{-\beta/2 (U(x)-U(x_i)} ~\int_{x(t_i)=x_i}^{x(t)=x} \mathcal{D} x~ e^{-\beta~S_{eff}[x]},
\ee
where $S_{eff}[x]$ is the effective "action",
\be
\label{Seff}
S_{eff}[x] = \int_{0}^t d\tau~ \Big(\frac{\gamma}{4}~\dot{x}^2~ + V_{eff}(x)\Big).
\ee
The pre-factor $e^{-\beta/2 (U(x)-U(x_i)}$ in Eq. (\ref{G}) can be transformed away, noticing that $\frac{d U(x)}{d \tau}= \dot x U'(x)$. One than obtains a path integral in which the statistical weight contains 
the  Onsager-Machlup functional
\be
\label{OM}
G(x,t|x_i) &=& \int_{x(0)=x_i}^{x(t)=x} \mathcal{D} x~ e^{-\beta~\int_{0}^t d\tau~ \Big(\frac{\gamma}{4}~\dot{x}^2~ + \frac12~\dot x ~U'(x)+ V_{eff}(x)\Big).}.
\ee

Eq. (\ref{GPI}) provides an expression for the conditional probability in terms of the microscopic stochastic dynamics governing the system. It represents the starting point of the Dominant Reaction Pathway 
approach~\cite{DRP1,DRP2,DRP3,DRP4}, which deals with the problem of finding the most probable transition pathways between the given configurations $x_i$ and $x_f$, which are visited at the initial and final time 
$x(t)=x_f, x(0)=x_i$, respectively.

On the other hand, in this work we are interested in the corresponding initial value problem, i.e. we are want to  develop an effective theory which yields directly the long-time evolution of the probability
density $P(x,t)$,  solution of Eq. (\ref{FP}), starting from a given initial probability density $P(x,t=0)=\rho_0(x)$.    The  probability density $P(x,t)$, the Green's function $G(x_f,t|x_i,t_i)$ and the initial distribution $\rho_0(x)$ 
are related by the Eq.
\be
P(x,t) = \int d y ~G(x,t|y)~\rho_0(y).
\ee 
Hence, for positive time intervals, the conditional probability $G(x,t|x_i)$ can be considered as the propagator associated to the stochastic Fokker-Planck Eq.~(\ref{FP}).
\section{Separation of Fast and Slow Modes}
\label{separation}

Without loss of generality, let us focus on the stochastic path integral (\ref{GPI}),  with periodic boundary conditions:
\be
\label{GPIp}
 Z(t) \equiv \int dx\ G(x,t|x,0) = \oint\ \mathcal{D} x~\exp\left[-\beta ~\int_0^t d \tau\Big( ~\frac{\gamma}{4} \dot x^2 + V_{eff}(x)\Big) \right].
\ee
We observe that  the inverse temperature $\frac{1}{\beta}$ plays the role of $\hbar$,  in the analogy with the quantum mechanical formalism. Hence, the loop expansion of the path integral (\ref{GPIp}) generates an expansion in powers of $\frac 1 \beta$. 

Let us introduce the Fourier conjugate:
\be
\x(\omega_n) &=& \frac{1}{t}~\int_0^t d\tau \exp\left[- i \omega_n t \right] ~x(\tau) \\
x(\tau) &=& x(\tau+t) =  \sum_n \x(\om_n)  \exp\left[ i \omega_n t \right].
\ee
where $\om_n$ are the Matsubara frequencies:
\be
\omega_n = \frac{2 \pi}{t}~n, \qquad  n=0, \pm 1, \pm 2, \ldots.  
\ee

In numerical simulations, the integration of the over-damped Langevin Eq. is performed by choosing a finite elementary time step $\Delta t$. In frequency space, this implies the
existence of an ultra-violet cut-off $\Omega$, which is inversely proportional to $\Delta t$:
\be
\Om  \sim \frac{2 \pi}{\Delta t}.
\ee
Such a relationship becomes a strict equality in the case of periodic boundary conditions, as in Eq.~(\ref{GPIp}). In general, when the boundary conditions are not periodic, it represents
just an order-of-magnitude estimate of the largest Fourier  frequencies, which are  associated to a given choice of the integration time step $\Delta t$. 

Let us now introduce a parameter $0<b<1$ and split the frequency interval $(0,\Om)$ as $(0, b~\Om)~\cup~(b~\Om, \Om)$. 
Then the Fourier decomposition of a path contributing to (\ref{GPIp})) can be split as:
 \be
x(t)= x_>(t) + x_<(t),
 \ee
 where $x_<(t)$ and $x_>(t)$ will be referred to as the \emph{slow}- and \emph{fast}- modes respectively:
 \be
 x_<(t) &=& \sum_{|\om_n| \le b\Om}~\x(\omega_n) ~e^{~i \om_n t }\\
 x_>(t) &=& \sum_{b\Om\le |\om_n| \le \Om}~\x(\omega_n) ~e^{~i \om_n t }.
  \ee

The main purpose of this work is to develop a perturbation series  to systematically integrate out from the path integral the modes  with frequencies $\omega_n > b \Om$.
To this end, we begin by re-writing the "kinetic" term which appears in the effective action (\ref{Seff}) of the path integral (\ref{GPI}) as a sum of the kinetic energy of  slow and fast modes:
\be
\frac{\gamma}{4 }~\int_0^t d\tau ~\dot x^2 &=&~\frac{t~\gamma}{4 }~ \sum_{|\om_n|\le \Om} \om_n^2~ \x(\om_n)~ \x(-\om_n)\nonumber\\
&=& \frac{\gamma}{4}~\int_0^t d\tau ~\dot x^2_<(\tau) + ~\frac{\gamma~t}{4 }~ \sum_{|\om_n|\in S_b} \om_n^2~ \x(\om_n)~ \x(-\om_n),
\ee
where $S_b$ denotes the shell of hard modes $S_b = \Sb$.
 
Let us now consider the potential term and expand around the slow modes $x_<(t)$
\be
~V_{eff}(x(\tau)) &=&V_{eff}(x_<(\tau)) +\frac{\partial V_{eff}(x_<(\tau))}{\partial x^i}~ x^i_>(\tau) + \frac 12  
~\frac{\partial^2 V_{eff}(x_<(\tau))}{\partial x^i x^j}~ x^i_>(\tau)~x^j_>(\tau)  + \mathcal{O}(x_>^3)
\ee
The complete path integral  (\ref{GPIp}) can be split in the following way:
\be
\label{effAct}
Z(t)~&=&  \oint   \mathcal{D}x_<\oint \mathcal{D}x_>~ e^{-\beta~S_{eff}[x_<(t)+x_>(t)]}\nonumber\\
&\equiv& \oint  \mathcal{D}x_<~ e^{-\beta~S_{eff}[x_<(t)]}~e^{-\beta S_>[x_<(\tau)]}.
\ee
in this expression, the action  functional $S_{eff}$ is evaluated on the slow-modes only and depends  on the original effective potential $V_{eff}$ (which we also shall refer to as the "tree-level" effective potential). 
$S_>[x_<(\tau)]$ is a correction term action which accounts for the
dynamics of the fast modes which are integrated out: 
\be
e^{-\beta S_>[x<(\tau)]} \equiv \oint  \mathcal{D} x_>~
e^{-\frac{\beta ~\gamma~ t}{4}~\sum_{|\om| \in S_b}~ \om_n^2 ~\x(\om_n)~ \x(-\om_n) - \beta S_{int}},
\ee
where the $S_{int}$ is an effective interaction term.
In such an Eq., the  integration over the hard modes is performed in Fourier space,
\be
\mathcal{D} x_> \equiv \prod_{|\om_n|~\in S_b} d \x(\om_n).
\ee
Eq. (\ref{effAct}) is formally exact. In the next section, we evaluate the effective action $S_{>}[x_<(\tau)]$ perturbatively. The effective interaction which includes the correction coming from $S_>[x_<]$ will be
referred to as the \emph{renormalized} effective interaction. 
 
\section{Renormalized Effective Interaction} 
\label{renormalized}

\begin{figure}[t]
\includegraphics[width=14 cm]{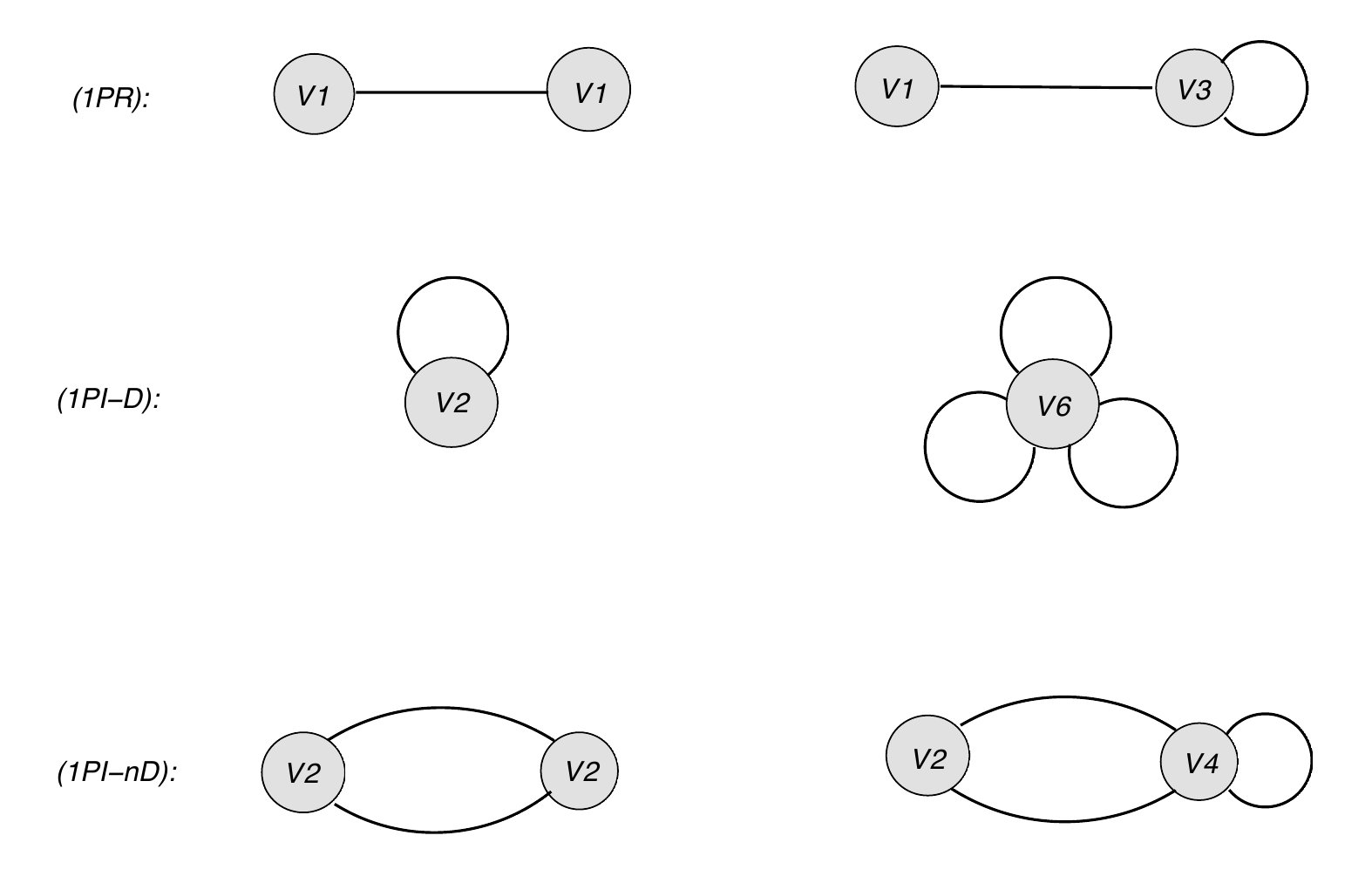}
\caption{Examples of connected graphs appearing in the exponent of
  Eq. (\ref{exact}). The diagrams
 on the upper part (dumbbell diagrams) are one-particle reducible, 
while those in the middle and in the bottom are one-particle-irreducible. In particular,
those in the middle (daisy diagram) are local in time.}
\label{types}
\end{figure}

In the previous section, we have seen that the the integration over the fast modes generates an additional term in the effective action for the slow modes:
\be
Z(t)~\equiv  \oint  \mathcal{D}x_<~ e^{-\beta~S_{eff}[x_<(t)]]-\beta S_>[x_<(\tau)]}
\ee
where
\be
e^{-\beta ~ S_>[x<(\tau)]} = \oint \mathcal{D}x_> ~e^{-\beta t~\sum_{|\om_n|\in S_b}~\frac{\gamma~\om_n^2}{4} ~\x(\om_n)~\x(-\om_n)}~e^{-\beta S_{int}} 
\ee

In this section we formally perform such an integration. We begin by
re-writing $e^{-\beta ~ S_>[x_<(\tau)]} $ as 
\be
e^{-\beta ~ S_>[x_<(\tau)]} = \langle e^{-\beta S_{int}} \rangle_0,
\ee
where the notation $\langle \cdot \rangle_0$ denotes the expectation
value evaluated in the free theory 
\be
S_>^0[x_>] &=& ~t~\sum_{|\om_n| \in S_b}~ \frac{\gamma ~\om_n^2}{4} ~\x(\om_n)~ \x(-\om_n).
\ee

To evaluate the matrix element $\langle e^{-\beta S_{int}}\rangle$, we represent the  $e^{-\beta S_{int}}$ ``operator'' by its power series:
\be
\label{sum}
\langle e^{-\beta S_{int}}\rangle_0 = \sum_k \frac{1}{k!} \langle (-\beta S_{int})^k \rangle_0 = \frac{1}{Z_>^0} ~\int \mathcal{D} x_> \sum_k \frac{1}{k!}~(-\beta S_{int})^k ~e^{-\beta S^0_>}.
\ee
Next, we expand the interaction action $S_{int}[x_>+x_<]$ around the slow modes\footnote{Throughout all this work, we shall adopt Einstein notation, i.e. the summation over repeated indexes
is implicitly assumed.}:
\be
\label{pert}
-\beta ~S_{int}[x_>+x_<] &=& - \beta \int_0^t d\tau ~\frac{\partial V_{eff}(x_<(\tau))}{\partial x^i}~x_>(\tau) 
-\beta~\frac 12  ~\int_0^t d\tau ~\frac{\partial^2 V_{eff}(x_<(\tau))}{\partial x^i x^j} x^i_>(\tau) x^j_>(\tau)+ \ldots \nonumber\\
&=& ~-\beta\int_0^t d\tau ~\sum_k ~\frac{1}{k!}~ V_{i_1, \ldots, i_k}(\tau) ~x^{i_1}_>(\tau)\ldots ~x^{i_k}_>(\tau),  
\ee
where 
$
 V_{i_1, \ldots, i_k}(\tau) ~x^{i_1}_>(\tau)\ldots ~x^{i_k}_>(\tau)
$ 
are vertices with couplings 
\be
\label{vertexes}
V_{i_1, \ldots, i_k}(\tau)  \equiv \frac{\partial^k V_{eff}[x_<(\tau)]}{\partial x^{i_1} \ldots \partial x^{i_k}}.
\ee
Notice that each term in the perturbative  expansion (\ref{pert}) generates a new vertex,   with an increasing power of the  $x_>(\tau)$ field. The  couplings to the fast modes 
depend implicitly on the time $\tau$,  through the slow modes $x_<(\tau)$. 

By Wick theorem, each term in the series (\ref{sum}) can be related to
a Feynman graph with vertexes given by (\ref{vertexes}) and
propagators given by ---see appendix \ref{propagator} ---: 
\be
\langle x^i_>(\tau_1)~x^j_>(\tau_2) \rangle_0 = \sum_{|\om_m|,|\om_n| \in
  S_b}~G^{0~i j}_>(\omega_n, \omega_m)~e^{i(\om_m\tau_1+\om_n\tau_2)}
=\sum_{|\om_n| \in S_b}~\delta_{i j}~\frac{2}{\beta ~\gamma ~t~
  \om_n^2}~e^{i\om_m(\tau_2-\tau_1)}.
\ee

The expansion~(\ref{sum}) can be re-organized as the exponent of the sum performed
over only connected diagrams: 
\be
e^{-\beta ~ S_>[x<(\tau)]} = e^{\textrm{(sum over all  connected diagrams)}}.
\ee
Hence, the path integral (\ref{effAct}) for the slow modes can be
given the following exact diagrammatic representation 
\be
\label{exact}
Z(t)~\equiv  \oint  \mathcal{D}x_<~ e^{-\beta~S_{eff}[x_<(t)]] +  \textrm{(sum over all  connected diagrams)}}.
\ee
Below we give a classification of all the connected diagrams that may
give a contribution to the expansion above. Firstly note that all
diagrams that involve an odd numbers of fast field vanish thanks to
the Wick theorem. We are thus left with the following sets of, a
priori nonvanishing, diagrams:\\
\begin{itemize}
\item 1PR (one-particle-reducible) diagrams, namely diagrams that can be
topologically separated into two distinct subdiagrams by cutting one
internal fast mode line (propagator): they have the topology of a
dumbbell. The simplest examples 
of dumbbell diagrams are depicted in the upper part
of Fig.~\ref{types}.  
\begin{figure}[t]
\includegraphics[width=18 cm]{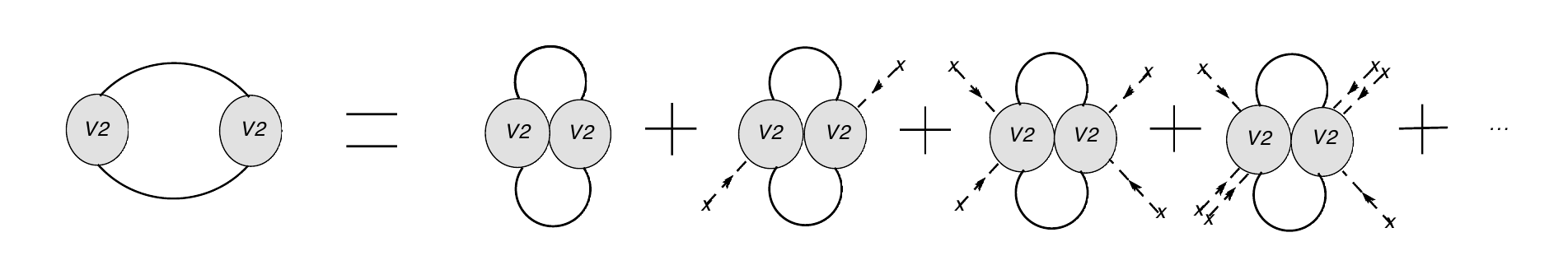}
\caption{Diagrammatic representation of the local time-derivative expansion of a non-local diagram ---Eq. (\ref{HD})---. Solid lines are fast-mode propagators, while dashed lines represent a single time derivative
acting on the corresponding vertex function.}
\label{flowers}
\end{figure}

The main assumption of this work is 
the existence of a gap between slow modes and fast modes. 
Under such assumption all the 1PR
diagrams give vanishing contributions.
From the physical point of view, this can be understood as a consequence of energy conservation: in order for the total energy flowing through  a vertex with a single hard mode to be conserved, at least one of the external modes has to be hard. On the other hand,  our working assumption implies that all the modes in the external legs of diagrams are soft.
This result can be rigorously proven for all 1PR.  As an example,  
we explicitly  compute the upper left diagram of
Fig.~\ref{types}. We have
\be
\frac{1}{2!}~\left(-\frac{\beta}{2!}\right)^2 ~\int_0^t d\tau_1~\int_0^t d\tau_2
~V_{i}~(x_<(\tau_1))~V_{j}~(x_<(\tau_2))
~\sum_{|\omega_n|\in S_b}~\frac{2}{\gamma~\beta~  t}~\frac{e^{i\om_n (\tau_2-\tau_1)}}{\omega_n^2}~~\delta_{i j}.
\label{eq:dumbbell}
\ee
We note that the effective potentials depend smoothly on time, through the periodic functions
$x^i_{<}(\tau)$. Hence,  the terms  $V_{i}[x_<(\tau_1)]$ and  $V_{j}[x_<(\tau_2)]$ in Eq.(\ref{eq:dumbbell}) can  be expressed in terms of their Fourier-transform,
\be
V_i(x_<(\tau_1)) &=&  \sum_n \tilde V_i(\nu_n)~ e^{i \nu_n \tau_1}\\
V_j(x_<(\tau_2)) &=&  \sum_m \tilde V_j(\nu_m)~ e^{i \nu_m \tau_2}.
\ee
This allows to perform the  time integrals, which  simply yield $t^2~\delta_{\om_n+\nu_n,0}~\delta_{\om_n-\nu_m,0}$.
Due to such delta-functions, only hard $\nu$-modes survives, which  are projected in a term 
\be
\propto~\sum_{|\nu_n|\in S_b}
 \frac{\beta t}{4\gamma ~ \nu_n^2}
\frac{1}{\nu_n^2}~\tilde V_i(\nu_n) \tilde V_i(-\nu_n) \approx 0.
\ee
These modes thus yield negligible contributions under the
 physical assumption of large separation of frequency scales.    On the other hand, if one does not assume a separation of time scale, this diagram gives finite contribution and has to be accounted for.
Note that this term has the same structure as the first correction which appears when one performs higher-order Trotter expansion\cite{trotter}. 

It is not difficult to check that such result holds for all 1PR
diagrams, so that we can reduce our effective action to the sum of
1PI (one-particle-irreducible) diagrams, i.e. diagrams that cannot be
 disconnected by cutting a single internal line. They can be
classified in two main groups:\\[2mm]
\item 1PI  ``daisy'' diagrams, namely diagrams with a single
vertex. Such diagrams only involve equal-time hard propagators and only give
rise to  contributions to the renormalized effective action which are local in time: they have the topology of a
daisy, hence the name. Examples of daisy
diagrams are depicted in the middle part of Fig.~\ref{types}. It is
not difficult to compute a generic daisy diagram with $K$ petals
(propagators). It is due to the vertex with $2K$ hard fields and reads
\be
(2K-1)!!~ \left(-\frac{\beta}{(2K)!}\right)\int_0^td\tau~~
\Delta^K V_{eff}(x_<(\tau))
\left(\frac2{\beta\gamma~t}\sum_{|\om_n|\in S_b}\frac1{\om_n^2}\right)^K,
\ee
where $\Delta \equiv \delta_{ij}~\partial_i \partial_j$ is the Laplacian operator, and the
numerical factor in front is a combinatorial factor. The sum, i.e. the
equal-time propagator, can be
easily performed by taking the continuum limit $\sum_\om \to
\frac{t}{2\pi}\int d\om$ that simply yields~\footnote{Here for later use we consider
a generic even power $2p$. It is easy to check that the error one
makes in considering the continuum limit is of order $\frac1{(1-b)N}$
with $N\equiv \Omega t/2\pi$. }
\be
\frac1t\sum_{|\om_n|\in S_b}\frac1{\omega_n^{2p}}\to
\frac1\pi\int_{b\Omega}^\Omega \frac{d\omega}{\omega^{2p}} =
\frac1{(2p-1)\pi}\frac{1-b^{2p-1}}{(b\Omega)^{2p-1}}, 
\label{sum-cont-limit}
\ee
so that we finally obtain
\be
\left(-\frac{\beta}{K!}\right)\left(\frac{D}\pi
\frac{1-b}{b\Omega}\right)^K \int_0^td\tau~~\Delta^K V_{eff}(x_<(\tau)),
\label{1PI}
\ee 
where we have reinstated the diffusion coefficient
$D=1/\beta\gamma$. Hence, one can even formally resum all the daisy
diagrams into the compact expression
\be
\sum {\rm daisy\ diagrams} = -\beta\int_0^td\tau~\exp\left({\frac
  D\pi\frac{1-b}{b\Omega}\Delta}\right) ~V_{eff}(x_<(\tau))~.
\label{sum-daisy}
\ee
\\[2mm]
\item 1PI non-daisy diagrams: all other non-local diagrams. The
simplest examples of such diagrams are depicted in the lower part of Fig.~\ref{types}. 
These diagrams generate contributions to the renormalized effective
action that are non-local in time and give rise to infinite series of
local diagrams. For example, the evaluation of the lower left diagram of
Fig.~\ref{types} yields a contribution of the form: 
\be
\label{spa}
 2\times \frac{1}{2!}~\left(-\frac{\beta}{2!}\right)^2 ~\int_0^t d\tau_1~\int_0^t
  d\tau_2
~V_{ij}~(x_<(\tau_1))~V_{kl}~(x_<(\tau_2))
~\sum_{|\omega_n|,|\om_m|\in S_b}~\left(\frac{2}{\gamma~\beta~
    t}\right)^2~\frac{e^{i(\om_n+\om_m) (\tau_2-\tau_1)}}{\omega_n^2
    ~\omega_m^2}~~\delta_{i k}~\delta_{j l}, \nonumber\\
\ee
where the $2$ in front is a combinatorial factor. After Fourier
transforming the potentials (see discussion below
eq.~(\ref{eq:dumbbell})), the integrals over times yield 
$t^2~\delta_{\om_m+\om_n+\nu_n,0}~\delta_{\om_m+\om_n-\nu_m,0}$.
Hence,
\be
\frac1{\gamma^2}\sum_{\nu_n}\sum_{|\omega_n|\in S_b}~\tilde
V_{ij}(\nu_n)~\tilde V_{ij}(-\nu_n)~\frac1{\omega_n^2}~\frac1{(\omega_n-\nu_n)^2}~.
\label{NL1}
\ee 

Now we again make use of the assumption that slow modes and
fast modes of physical processes under study are separated by a
large gap. Under such assumption we can safely expand the second
fraction in the latter expression in power series of slow modes
$\nu_n$ and rewrite~(\ref{NL1}) as higher-time-derivative
expansion. Let us reintroduce the integral over time as $1
=\frac1t~\int_0^td\tau~\sum_{\nu_m} e^{i(\nu_n+\nu_m)\tau}$ so that
powers of $\nu_n$ can be traded with time-derivative of the
potential (note that odd powers vanish upon symmetric sum; in fact they would
 give rise to total time-derivative terms that are zero upon
 integration thanks to periodicity in time.) We are thus left with
\be
\frac1{\gamma^2}\int_0^td\tau~V_{ij}(x_<(\tau)) \Biggl[\frac1t
  \sum_{|\omega_n|\in S_b}\frac1{\om_n^4} +\frac3t
  \sum_{|\omega_n|\in S_b}\frac1{\om_n^6}~(-\partial_\tau^2)
+\frac5t
  \sum_{|\omega_n|\in S_b}\frac1{\om_n^8}~\partial_\tau^4+\cdots\Biggr] 
~V_{ij}(x_<(\tau))~.
\ee      
Sums over hard frequencies can be performed in the continuum limit
with the help of formula~(\ref{sum-cont-limit}) and time-derivatives
can be partially integrated in order to rewrite the latter in a more
symmetric form
\be
\frac1{\pi~\gamma^2}\int_0^td\tau~\Biggl[\frac13\frac{1-b^3}{(b\Omega)^3}\Big(V_{ij}(x_<)\Big)^2 
+\frac35\frac{1-b^5}{(b\Omega)^5}\Big(V_{ijl}(x_<) ~\dot x^k_<(\tau)\Big)^2
+\frac57\frac{1-b^7}{(b\Omega)^7}\Big(V_{ijlm}(x_<) ~\dot x^l_<(\tau)~ \dot x^m_<(\tau)\Big)^2
+\cdots\Biggr]~.
\label{HD}
\ee
The infinite higher-derivative expansion is the legacy of non-locality
in time: such an expansion can diagrammatically represented as an
infinite sum of local (daisy-like) diagrams, as depicted in Fig.~\ref{flowers}. 
\end{itemize}
It is intuitive to expect that, in  the presence of decoupling low and high frequency modes, the higher-derivative terms should be
suppressed. In the next section, we shall generalize this statement and present a quantitative method to systematically organize all contributions to the effective action in terms of a perturbative series.

\section{Slow-mode perturbation theory}
\label{perturbative}

The diagrammatic representation of the path integral given by
Eq. (\ref{exact}) is formally exact, but rather useless. In fact,  it
is obviously impossible to evaluate and re-sum exactly all the
infinitely many Feynmann graphs  
appearing in the exponent. 
On the other hand, in this section we show that it is possible to
compute the renormalized effective action $~S_{eff}[x_<(t)]] $  
to an  \emph{arbitrary level of precision}, by calculating only a \emph{finite number} of Feynmann graphs. This way, the low-frequency effective theory retains predictive power.

The idea is to exploit the decoupling of the short-time dynamics from the long-time dynamics to  organize the sum over all possible graphs as a perturbative expansion.
  We shall refer to such a systematic evaluation of the renormalized low-frequency effective action as to the \emph{slow-mode perturbation theory}.

The first step in the construction of our perturbation series is to identify all the dimensionless combinations of the physical quantities which appear in the Feynmann graphs contributing to 
(\ref{exact}), evaluated in stationary phase approximation.  Let us
first define the quantities
\be
\bar V \equiv ~\frac1t \left| \int_0^td\tau~V\right | ~,
\qquad
\bar V^k_{2m}~\equiv~\frac1{\bar V} \left| \int_0^td\tau~\Delta^m V\right |
\sim k^{2m}~,\qquad
\bar V^\om_{2m}~\equiv~\frac1{\bar V^2} \int_0^td\tau~\Big(\partial_\tau^m V\Big)^2
\sim \om_<^{2m},
\ee
where $k$ is the typical wave vector on the spatial Fourier transform
of $V_{eff}(x)$ and $\omega_<$ is the typical frequency in temporal
Fourier transform of $V_{eff}(x_<(\tau))$.

Using these combinations, we can thus construct the following 
dimensionless combinations:
\be
\alpha_1 \equiv \frac{\beta \bar V}{b\Omega} ~,\qquad
\alpha_2 \equiv \frac{k^2D}{b\Omega} ~,\qquad
\alpha_3 \equiv \frac{\omega_<}{b\Omega}~. 
\ee
We are interested in describing the dynamics of physical systems for which each of these parameters can be considered small. 
In order to illustrate the physical interpretation of the condition $\alpha_1\ll 1$, we observe that the probability for the system to remain in the same configuration $x$,
 during an elementary time interval $dt$ is
\be
P(x,t+dt|x,t) \propto \frac{1}{(dt)^{d/2}}~e^{-\beta V_{eff}(x) dt}.
\ee
Hence, the combination $\beta V_{eff}$ represents\footnote{Notice that, in the small 
temperature limit, $V_{eff}(x)$ becomes positive definite. Thus, $P(x,t+dt|x,t)$ decays exponentially with $\beta V_{eff}$ in the time interval $dt$.} 
the typical time scale associated to local conformational changes, and the condition $\alpha_1\ll1$ expresses the condition that
  the time spent on average by the system in each configuration is large compared
to the elementary short-time scale, $dt\sim \frac{1}{b\Om}$.
\begin{figure}[t]
\includegraphics[width=14 cm]{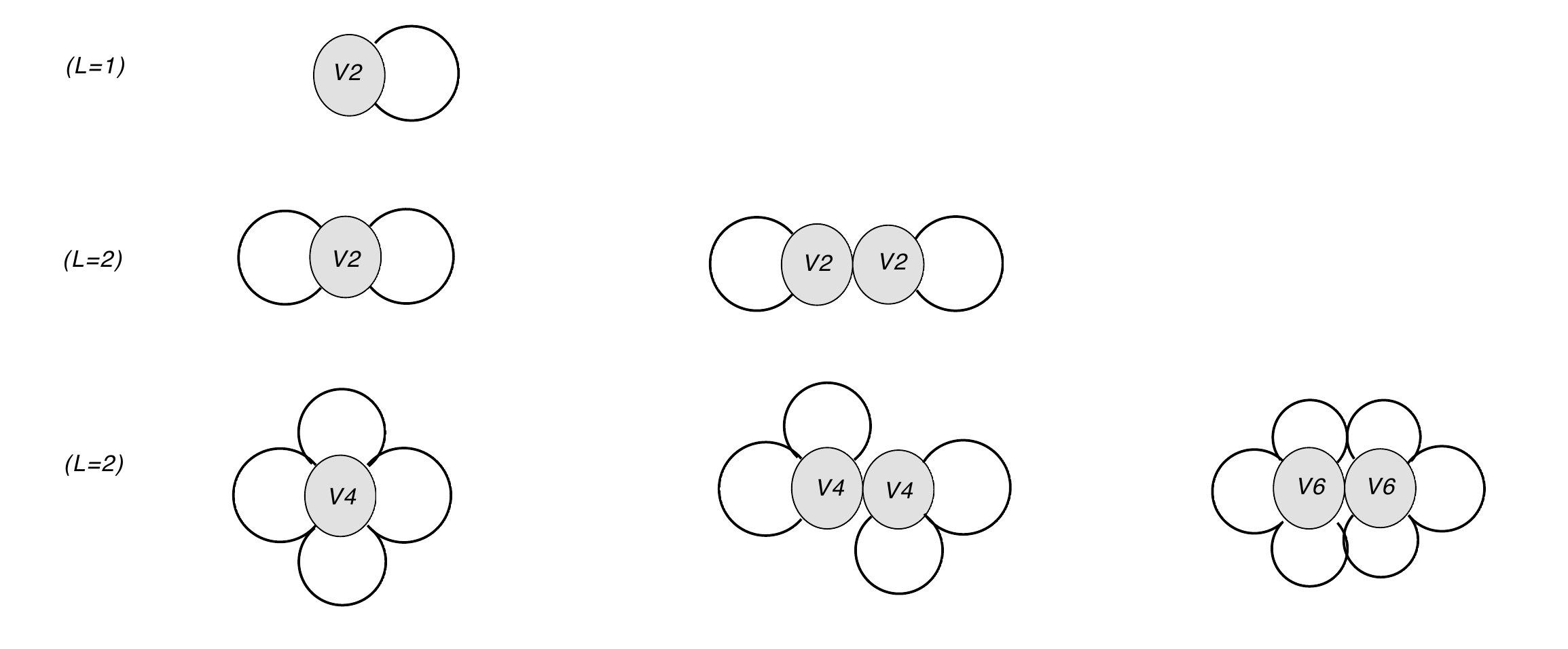}
\caption{Examples of the diagrams with the lowest degree of slowness, up to $L=2$. }
\label{Lfig}
\end{figure}

The condition $\alpha_2\ll1$ implies that the effective potential 
varies over length scales which are large, compared with the  mean distance covered by Brownian motion in an elementary time interval $dt$.
Finally, the condition $\alpha_3\ll 1$ implies that the typical slow mode frequencies are  small compared to the ultra-violet cut-off, which is of the order of the inverse of the elementary time interval $dt$.

It is easy to see that any local diagram in the expansion of the
renormalized effective action comes about with integer powers of these
coefficients, when compared to the tree level
effective action. In particular, any diagram composed by $r$ vertices
of $M_1,\ldots,M_r$ hard fields will involve $M=\sum_{i=1}^r M_i$
spatial derivatives and $\frac M2$ propagators each of which yields a
power of $\frac1{b\Omega}$. Finally, each additional vertex 
yields a power of $\beta \bar V$ and each
time derivative yields a power of $\omega$. So, the above diagram, at
the lowest level in time derivatives will be of order $ \alpha_1^{r-1}~\alpha_2^{\frac M2}$
with respect to the tree level expression. Higher time derivative
terms will add powers $\alpha_3$. It is thus natural to define a
{\em degree of slowness} $L$ for a local diagram, given by
\be
\label{slow}
\vspace{2 cm}\nonumber\\
L({\rm Feynman\ diagram}) = N_v-1 +N_\tau+\frac{N_x}2,\\
\vspace{2 cm}\nonumber
\ee    
where $N_v$ is the number of vertices, $N_\tau$ the number of time
derivatives and $N_x$ the number of spatial derivatives. The definition in Eq. (\ref{slow}) is
normalized in such a way that $L({\rm tree}) =0$. 

It is easy to check that the degree of slowness $L$  corresponds to the power of $\frac1{b\Omega}$ of the local diagrams. Note also
that for daisy diagrams and for all other diagrams where
$N_\tau=0$, the degree $L$ is nothing but the number of loops. One can
thus easily write down and compute the finite set of local diagrams
that renormalize the effective action up to a fixed (yet arbitrary) level of
precision $L_{max}$. Let us consider a few simple examples. 
\begin{itemize}
\item $L\leq 1$
corresponds to a single daisy diagram with $L=1$, $N_v=1$ and $N_x=2$, represented in the left panel of Fig.\ref{Lfig}. The expression of this diagram is given by Eq.~(\ref{1PI}) and gives a correction
 to the effective action of the form
\be
S_{>}[x_<;L\leq 1]=\frac{D}\pi
\frac{1-b}{b\Omega} \int_0^td\tau~~\Delta V_{eff}(x_<(\tau)).
\label{L=1}
\ee   

\item $L\leq 2$ corresponds to two further diagrams, one daisy diagrams with
either $N_x=4$ and the two-vertex local diagram with $N_x=2$ and no
time-derivatives. This latter however is 1PR and gives no
contribution. We are thus left with the corrections 
\be
S_{>}[x_<;L\leq 2]=\frac{D}\pi
\frac{1-b}{b\Omega} \int_0^td\tau~~\Delta V_{eff}(x_<(\tau))
+\frac{1}{2}\left(\frac{D}\pi
\frac{1-b}{b\Omega}\right)^2 \int_0^td\tau~~\Delta^2 V_{eff}(x_<(\tau))
\label{L-leq-2}
\ee   

\item $L\leq 3$ corresponds to two further diagrams, one daisy diagrams with
$N_x=6$ and the two-vertex local diagram with $N_x=4$ and no
time-derivatives. This latter can be simply read off from
eq.~(\ref{HD}). Hence
\be
S_{>}[x_<;L\leq 3]&=&\frac{D}\pi
\frac{1-b}{b\Omega} \int_0^td\tau~~\Delta V_{eff}(x_<(\tau))
+\frac{1}{2!}\left(\frac{D}\pi
\frac{1-b}{b\Omega}\right)^2 \int_0^td\tau~~\Delta^2 V_{eff}(x_<(\tau))
\nonumber\\
&&\frac{1}{3!}\left(\frac{D}\pi
\frac{1-b}{b\Omega}\right)^3 \int_0^td\tau~~\Delta^3 V_{eff}(x_<(\tau))
-\frac{\beta~D^2}{3\pi}\frac{1-b^3}{(b\Omega)^3}\int_0^td\tau~
\Big(\partial_i\partial_jV_{eff}(x_<(\tau)) \Big)^2 
\label{L-leq-3}
\ee   
that involves in the last term the trace of the square of Hessian of
the tree level potential $(\partial_i\partial_jV(x_<)
)^2 ={\rm Tr}~ {\cal H}^2_V$. In order to see the first time
derivatives appearing into the renormalized effective action 
we need to consider $L\leq 5$ where, along with several other
corrections, we have the correction coming from the second term
in~(\ref{HD}) that yields
\be
-\frac{3\beta D^2}{5\pi}\frac{1-b^5}{(b\Omega)^5}\int_0^td\tau~{\rm
  Tr}~ {\cal \dot H}^2_V
\ee 
that can be also recast as a correction of the kinetic action
\be
\beta\gamma\int_0^td\tau~\Biggl[\frac14\delta_{kl} -\frac{3\beta^2
    D^3}{5\pi}\frac{1-b^5}{(b\Omega)^5}\partial_i\partial_j\partial_k
  V ~ \partial_i\partial_j\partial_l V\Biggr]\dot x^k \dot x^l~.  
\ee 
 \end{itemize}

Some comments on the results obtained in this section are in order.
First of all, we  emphasize  that  the effective interactions have been derived under the assumption that the modes which are relevant for the long-time dynamics 
 vary over time scales much longer than that of the fast modes, which enter in the loop diagrams. 
This is the crucial assumption of all renormalization group approaches. 
Our results confirm the intuitive picture  that if one adopts a low "time-resolution power", then the effective interactions generated by the ultra-violet modes can be regarded as 
instantaneous. This is in fact  general property of renormalization group theory, which is preserved to any order in the perturbative expansion.
Finally, we note that the correction terms generated by the
integration over the fast modes is suppressed, in the small
temperature limit.

\section{Renormalization Group Improved Monte Carlo}
\label{MC}

The usefulness of  the renormalization procedure resides in the fact that it gives rise to an effective theory, in which the largest frequency scale is lowered form $\Om$ to $b \Om$. Equivalently, the 
shortest time scale is increased form $\Delta t$ to $\frac{1}{b} ~\Delta t$. 
By construction, in the regime of decoupling of fast and slow modes, the probability density generated by the new slow-mode effective theory
must be the same as that of the original (i.e. tree-level) theory. In this section, we show how it is possible to use the slow-mode effective theory to develop improved MC algorithms for the time evolution of the probability density $P(x,t)$,
 in which the elementary time step used to propagate the configurations is increased by a factor $1/b$.

The starting point of the MC approach \cite{DMC} is to write the probability of observing the system in configuration $x$ at time $t$ in terms of the Green's function of the Fokker-Planck Eq. $G(x,t|x_i,t_i)$:
\be
\label{prob}
P(x,t) =  \int d x_i~ G(x,t|x_i,t_i)~ \rho_0(x_i)
\ee
where $\rho_0(y)$ is the density of states at the initial time.  

One then uses Trotter's formula to write the transition probability as a sequence of intermediate elementary propagation steps:
\be
P(x,t) =  \int  \prod_{k=0}^{N-1} ~d y_k ~G(y_{k+1},t_{k+1}|y_k,t_k) ~\rho_0(x_i)\qquad (y_0=x_i, y_N=x)
\ee
If  a sufficiently large number of intermediate steps  $N$ is adopted,  then the time steps $\Delta t = t_{k+1}-t_k$ can be considered infinitesimal and  the (unnormalized) transition probability  
$G(y_{k+1}, t_{k+1}| y_k,t_k) $ can be calculated analytically
\be
\label{product}
G(y+dy,t+\Delta t| y,t)  = \textrm{const.}~\times~e^{-\beta\left(\frac{ \gamma}{4}\left(\frac{dy}{\Delta t}\right)^2~\Delta t + \frac{1}{2} \frac{dy}{\Delta t}  \cdot \nabla U(y)\right)~\Delta t}
~e^{-\beta~ V_{eff}(y) \Delta t} 
 \ee
"Completing the square" in the first exponent, one finds
\be
G(y+dy,t+\Delta t| y,t)  = \textrm{const.}~\times~e^{-\frac{1}{4 D~\Delta t}~\left(  dy + \frac{\Delta t}{\gamma}~\nabla U(y)\right)^2 + \frac{\beta}{4\gamma }( \nabla U)^2\Delta t}~e^{ -\beta ~V_{eff}(y)~ \Delta t} 
 \ee
Now and recalling the definition of the effective potential (\ref{Veff}) in the second exponent,  this Green's function can be written as:
\be
\label{product2}
G(y+dy,t+\Delta t| y,t)  = \textrm{const.}~\times~e^{-\frac{1}{4 D~\Delta t}~\left(  dy + \frac{\Delta t}{\gamma}~\nabla U(y)\right)^2 }~e^{ \frac{1}{2 \gamma }\nabla^2 U(y)~ \Delta t} 
 \ee

In the  MC algorithm,  one starts from a set of initial system's configurations,  sampled according to he distribution $\rho_0(x_i)$. Such an ensemble is  evolved in time, 
according to the following procedure.
Each configuration is propagated for an elementary time interval $\Delta t$, by sampling from the Gaussian
\be
~e^{-\frac{1}{4 D~\Delta t}~\left(  dy + \frac{\Delta t}{\gamma}~\nabla U(y)\right)^2 }~
\ee
 in Eq.~(\ref{product2}).  
Such a configuration is then re-weighted according to the factor 
\be
\label{Wtree}
\mathcal{W}(y)=~e^{  \frac{1}{2\gamma}~ \nabla^2 U(y) \Delta t}. 
\ee
The iteration of such a procedure for many consecutive elementary propagations gives rise to a set of diffusive trajectories, called walkers. 
In the so-called \emph{diffusion} MC algorithm, the term $\mathcal{W}$ is used to replicate or annihilate the walkers. The ensemble of configurations obtained according to this procedure is distributed according to the probability density (\ref{prob}).

For the MC algorithm to be efficient, the fluctuations in the statistical weight of the walkers ---or, equivalently, in the number of walkers--- should remain small, throughout the entire time-evolution. 
This condition is verified if the factor $\mathcal{W}(y)$ is  always of order one. 
Note however that this term tends to enhance (suppress) the weight  of configurations in the vicinity of  the local minima (maxima) of $U(y)$, where the Laplacian is positive (negative). 
Hence, if the energy landscape varies very rapidly in space, then the fluctuations in the statistical weights ---or in the number of walkers--- will in general be large,  unless the elementary time step $\Delta t$ is chosen very small.
This feature represents a limiting factor of MC simulations, which makes the sampling of the probability density at large times very computationally  expensive. 
\label{example}
\begin{figure}[t]
\includegraphics[width=10 cm]{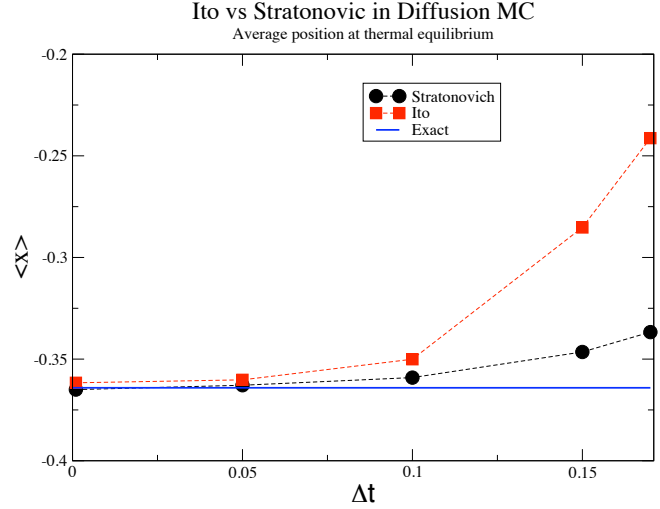}
\caption{Average position at thermal equilibrium, obtained from diffusion MC simulations with (circles) and without (square) the 
branching factor $\mathcal{W}(t)$ of Eq. (\ref{Wtree}), for different values of the discretization time step $\Delta t$.  Errors are smaller than the symbols.}
\label{itostrat}
\end{figure}

Clearly, the elementary propagation time step $\Delta t$ is the shortest time scale in the simulation. On the other hand, in the slow-mode effective theory one integrates out the dynamics in the  time scale range $(\Delta t,1/b \Delta t)$. Hence,
 we expect that by taking into account of the corrections associated to the renormalized effective interaction  it is possible to  perform MC simulations in which the 
 integration time step $\Delta t$ is chosen a factor $1/b$ larger. 

In practice,   the effective slow-mode theory introduces a correction in the re-weighting ---or branching---. To order $L=1$ one has
\be
\mathcal{W}_{L=1}(y) =~e^{  2 \nabla^2 U(y) \Delta t}  ~\times~e^{ -\frac{D}\pi
\frac{1-b}{b\Omega} ~\nabla^2 V_{eff}(y) \Delta t}
\ee
Notice that this expression contains a factor of the inverse frequency cut-off $1/\Omega$ in the  exponent. Such a term is proportional to 
the elementary time step $\Delta t$. The corresponding proportionality factor  reads $2\pi$ only for periodic path integral. For a generic initial value MC one can write in general
\be
\Om = \kappa ~\frac{2 \pi}{\Delta t},
\label{ka}
\ee
where the constant $\kappa$ is to be determined from simulations.
Hence, we obtain
\be
\label{eW}
\mathcal{W}_{L=1}(y) =~e^{  2 \nabla^2 U(y) \Delta t}  ~\times~e^{ -\kappa~\frac{D}{2 \pi^2}
\frac{1-b}{b} ~\nabla^2 V_{eff}(y) \Delta t^2}.
\label{WL1}
\ee

The unknown constant $\kappa$ can be determined by matching the results obtained by running a \emph{short} simulation in the  tree-level theory ---i.e. using an integration step $\Delta t$ and the tree-level weighting term (\ref{Wtree})--- with those obtained  in the effective
theory --- i.e. using an integration step $1/b~\Delta t$ and the renormalized weighting term  (\ref{eW})---. In the regime of decoupling of fast and slow modes, once the matching has been done, the two algorithms must generate the same evolution for the probability density at any later times.

In the next session, we shall provide an example which illustrates how this procedure works in practice and show that the fundamental and the effective theory do indeed generate the
same long-time stochastic dynamics.

\section{An illustrative Example}
\label{exa}

In order to illustrate how the renormalization of the effective interaction works in a simple example, let us consider the dynamics of a point particle, diffusing in a rugged 
asymmetric harmonic oscillator:
\be
\label{U2D}
U(x) =  h_1 x^2 + h_2 x + h_3 \sin(w x), 
\ee 
with $h_1=2,~ h_2=1,~ h_3= 1,~ w=4$. The   viscosity coefficient is set to $\gamma=5$  and inverse temperature to $\beta=5$.     Note that this potential has been chosen in such a way that the average value of $x$ at thermal equilibrium is non-vanishing.

The diffusion Monte Carlo algorithm used in our numerical simulations is presented in the appendix \ref{algo}. 
The factor $\Omega$, which appears  in the $L=1, L=2$ improvement terms was determined from the time interval $\Delta t$ using Eq. (\ref{ka}). The proportionality constant $\kappa$ in Eq.~(\ref{eW}) was determined once and for all, by matching the result of $\langle x(t) \rangle$ of the unimproved (i.e. $L=0$)  simulations after 10 integration time steps with $\Delta t=0.01$, with those of the RG-improved (i.e. $L=1, L=2$) MC simulations after a single elementary time step, with $\Delta t' = 0.1$. We found $\kappa=0.35$, with no appreciable difference between the $L=1$ and $L=2$ estimates.

Let us now discuss the results of our simulations. We begin by analyzing the effects of accounting for the factor $\mathcal{W}(x)$ defined in  Eq. (\ref{Wtree}), in  numerical MC simulations.
Fig.~\ref{itostrat} shows the average position, once the system has attained thermal equilibrium, obtained by diffusion MC simulations with and without branching the walkers according to $\mathcal{W}(x)$.  We recall that neglecting such a term is equivalent to simulating the dynamics in the Ito calculus, while the branching  is expected to improve the time discretization to order $\Delta t^2$. 
 Indeed, our results show that, when one chooses small discretization steps, the two approaches are consistent with each other and yield the exact equilibrium average --- which was computed directly from the Boltzmann distribution---. On the other hand, at large discretization steps, accounting for the factor $\mathcal{W}$ significantly improves the result.
The same discussion can be trivially repeated in simulations in which the factor $\mathcal{W}(x)$ is interpreted as a re-weighting term, while the number of walkers is held constant. 

\begin{figure}[t]
\includegraphics[width=21cm]{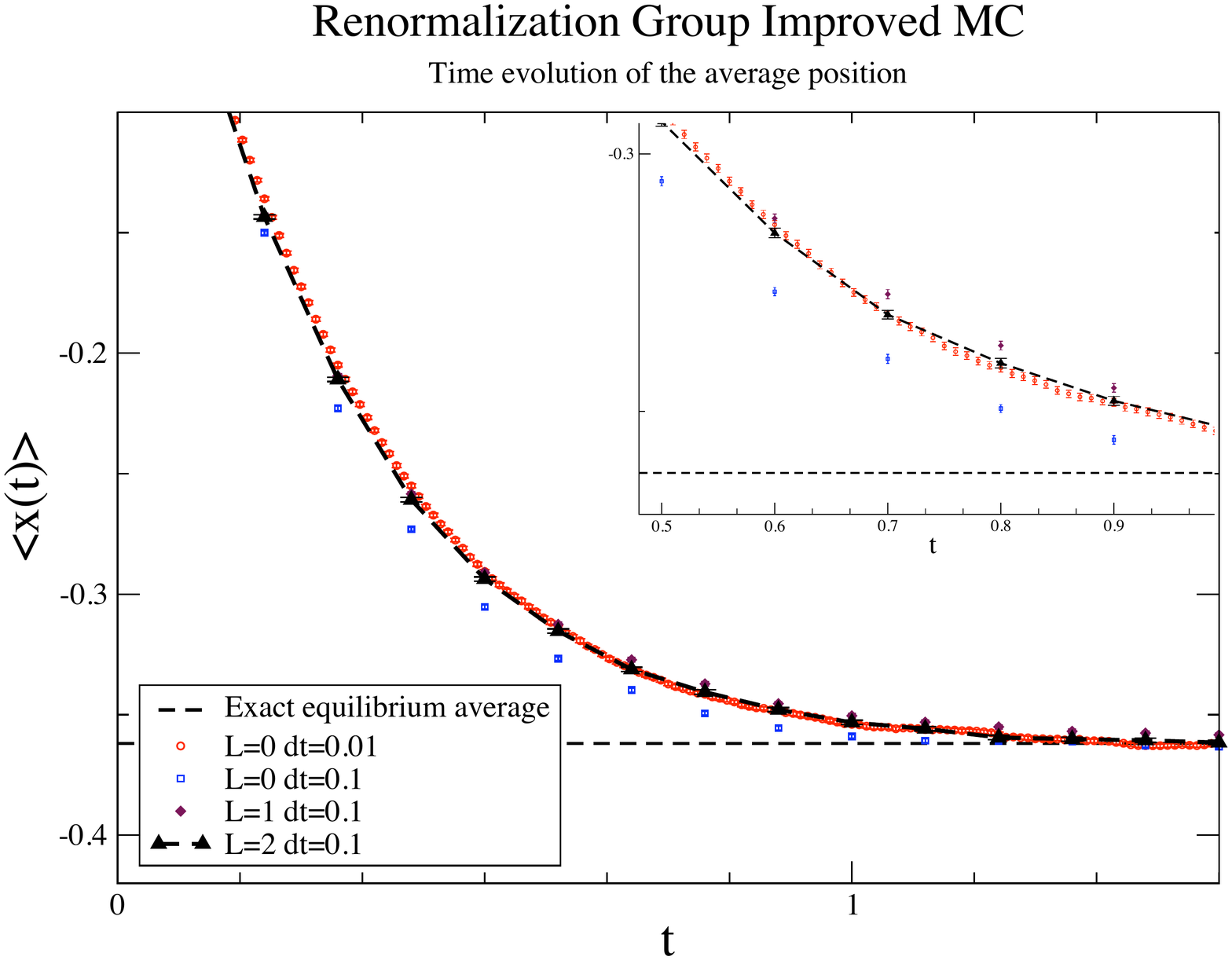}
\caption{ 
The average position of the particle at time $t$, computed in the tree-level theory ($\Delta t=0.01$ for $L=0$), and 
in the effective theory ($\Delta t=0.1$ for $L=1$ and $L=2$). The insert displays a part of the same curve,  on a larger scale. 
Statistical errors are smaller than the symbols. The $\Delta t=0.1$, $L=2$ cannot be distinguished from the $\Delta t=0.01$, $L=0$ curve. }
\label{Wfigs}
\end{figure}

 We now discuss the use of our effective theory to simulate the stochastic dynamics, using large time steps.
Fig.~\ref{Wfigs} shows the  time evolution of the average particle position  at time $t$, computed using a small discretization time step ---$\Delta t=0.01$--- and a large discretization step --- $\Delta t'=0.1$---. 
The two curves obtained in the original ---i.e. tree-level--- theory are compared with the results of the effective theory at order $L=1$ and $L=2$, which were obtained using an
 integration time step which was one order of magnitude larger,  $\Delta~t'=0.1$.

The time evolution of the observable $\langle x(t)\rangle$, obtained in the tree-level theory using  large integration time steps (squares) is  inconsistent with the
same quantity obtained using small time steps $\Delta t=0.01$ (circles).  This is expected, because for $\Delta t=0.1$ the numerical simulations of the tree-level theory start to be affected by significant discretization errors ---see Fig. \ref{itostrat}---.

The results of simulations with  large discretization time steps are significantly improved if one uses the effective theory, already at order $L=1$ (diamonds). 
At order $L=2$ the dynamics of the tree-level theory simulated at $\Delta t=0.01$ is indistinguishable from the dynamics of the effective theory simulated with $\Delta t'=0.1$ (triangles).
These results show that the
 hard-mode dynamics in the short time range from $0.01$ to $0.1$ has been correctly taken into account by means of the renormalized effective interaction. 
 As a consequence, the use of the effective theory allows to obtain  very accurate predictions, using larger time steps. 


\section{Long time dynamics of Molecular Systems}
\label{molecular}

The improvement of the MC algorithm based on our effective theory  is expected to be most efficient when the gap between the slow and the fast modes is very large. In fact, in this regime, 
the slow-mode perturbation theory remains reliable even when one
integrates out a large frequency shell, i.e. when $b\ll 1$. Hence, in this case, by RG-improvement it is possible  to simulate the time evolution using  elementary time steps $\Delta t'$ which are  significantly 
larger than the original elementary time step $\Delta t$, which would be used in the usual (unimproved) MC algorithm.

A natural application of the RG-improved MC is the investigation of the long-time dynamics of macromolecules, for which standard MD or MC algorithms can be extremely computationally expensive. Hence,  it is interesting to address the question of what is the typical range of reliability of the slow-mode 
perturbation theory  for a typical molecular interaction, at room temperature.
To this end, let us consider the over-damped diffusion at temperature $300~$K of   two molecules of mass $m=30~$amu, 
interacting through a Van-Der-Waals potential:
\be
\label{VdW}
U(r) = 4\epsilon \left[\left(\frac{\sigma}{r}\right)^{12} - \left(\frac{\sigma}{r}\right)^6 \right]
\ee   
where $\epsilon=4$~KJ/mol and $\sigma= 0.3$~nm. A typical value for the viscosity coefficient for a molecule in its solvent (e.g. an amino acid in water)
is $\gamma \sim~2\times 10^3$~amu~ps$^{-1}$.The typical time-steps used in the numerical 
integration of the Langevin Eq. (\ref{langevin}) are of the order $\Delta t\simeq 10^{-3}-10^{-2}$~ps. 
 
\begin{figure}[t]
\includegraphics[width=10 cm]{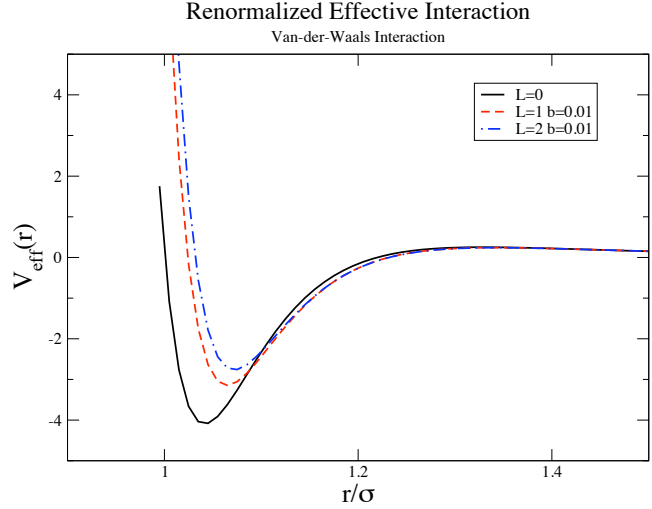}
\caption{ The tree-level, $L=1$  and $L=1$ renormalized effective potential $V_{eff}(r)$ for the Van-der-Waals interaction  Eq. (\ref{VdW}), 
 obtained integrating out the modes in the shell $S_b=(b \Omega, \Omega)$ with $\Omega = 2\pi /0.01 ps$ and $b=0.01$.}
\label{renormVdW}
\end{figure}

 The tree-level effective interaction associated to the potential (\ref{VdW}) is:
 \be
 V_{eff} (r) =  \frac{1}{4 \gamma} \left\{ \left[ 24 \epsilon ~\frac{\sigma^6}{r^7}~\left( 1 - 2\frac{\sigma^6}{r^6}\right) \right]^2
 - \frac{8 \epsilon}{\beta}~\frac{\sigma^6}{r^8}~\left[~156 \frac{\sigma^6}{r^6}-42\right]\right\}
 \ee
This function and the corresponding $L=1$ and $L=2$ renormalized effective interactions are plotted in Fig. \ref{renormVdW} for $\kappa=1$. 
This plot shows that, for a realistic set of parameters,
 the perturbative expansion remains reliable even when one integrates out a very large shell of modes, with  $b \sim 10^{-2}$.  
This fact suggests  that the ultra-violet dynamics is essentially free brownian motion, while the long time dynamics is dominated by very low-frequency modes,
and is driven by the force field.
This fact has remarkable consequences on practical numerical simulations. It implies that by using the renormalized effective potential, 
it should be possible to adopt integration time steps which are about $10^2$ time larger than those required to simulate the dynamics in the original tree-level theory. 
    

\section{Conclusions}
\label{conclusions}

In this work, we have presented a new approach to the problem of  investigating the long-time out-of-equilibrium dynamics of multi-dimensional systems obeying Langevin dynamics. 
In the presence of decoupling of time scales, the methods based on the direct integration of the Langevin Eq. (MD) or on the time propagation of the Fokker-Planck probability density (MC) 
are usually inefficient, because a significant amount of computational time in invested to simulate uninteresting fast stochastic fluctuations.

We have shown that the decoupling of time scales which limits MD and MC approach can in fact be exploited to  
perform analytically the average  over the short-time stochastic fluctuations. 
After the integration over the fast modes has been performed, one obtains an effective theory which describes directly the relevant dynamics, with a lower time resolution. 
In such an effective theory, the effective action in the path integral receives corrections, which account for the ultra-violet physics which is cut-off.
We have developed a rigorous  scheme which allows to organize such corrections in term of a perturbative series in which the expansion parameters are the ratio between the
soft frequency scales and the hard frequency scale $b \Omega$. Hence, sub-leading terms in the perturbative expansion come with higher inverse powers of the hard scales $b \Omega$ and become irrelevant in the limit in which the decoupling of 
fast and slow modes is very large.

The Feynmann diagrams which have to be calculated to obtain the corrections to any given order in this perturbation theory can be identified from their degree of slowness 
\be
L({\rm Feynman\ diagram}) = N_v-1 +N_\tau+\frac{N_x}2
\ee
Diagrams with degree of slowness $L$ generate corrections proportional to $1/(b\Om)^L$.
In particular, we have found that the leading-order correction (i.e. $L=1$) is proportional to the Laplacian of the effective potential $V_{eff}$:
\be
S_{>}[x_<] \simeq \frac{D}\pi
\frac{1-b}{b\Omega} \int_0^td\tau~~\Delta V_{eff}(x_<(\tau)).
\ee   
At the next-to-leading order, a term containing fourth-order derivatives appears:
\be
S_{>}[x_<] \simeq \frac{D}\pi
\frac{1-b}{b\Omega} \int_0^td\tau~~\Delta V_{eff}(x_<(\tau))
+\frac{1}{2}\left(\frac{D}\pi
\frac{1-b}{b\Omega}\right)^2 \int_0^td\tau~~\Delta^2 V_{eff}(x_<(\tau)).
\ee   
On the other hand,  a space-dependent, tensor 
correction to the diffusion coefficient  appears only as a higher-order effect ($ L=5$).
 It is important to stress the fact that, in the present approach, the ultraviolet cut-off $\Om$ (or, equivalently, the short time scale $\Delta t$) is kept finite at all stages. Upon taking the continuum limit $\Delta t\to 0$, all the correction terms in the effective theory vanish and one recovers the original theory, defined by the effective Schr\"odinger Eq. (\ref{SE}).

The main usefulness of such an effective theory resides in the fact that it can be used to develop an improved MC approach, to compute the long-time evolution of the Fokker-Planck probability.  The elementary time steps used in the RG improved MC algorithm are a factor $1/b$ larger those of the  MC algorithm for the underlying tree-level theory. Since the dynamics in the time range  $(\Delta t,1/b ~\Delta t)$ is averaged analytically, the RG improved MC algorithm 
 avoids investing  computational time in simulating the fast-mode dynamics associated to local Brownian motion. 
 
In the specific case of molecular interactions at room temperature, we have shown that the perturbative approach remains reliable even when one integrates large frequency shells, with $b\simeq 0.01$. 
This feature suggests that, by using the effective theory, it is possible to  simulate time intervals which can be up to a factor $\sim 100$ longer than in the usual MC approach.


\appendix

\section{Propagator of the fast modes}
\label{propagator}

Here we derive free fast mode propagator $G^0_>(\omega_n, \omega_m)$ appearing in the diagrams, using the standard source technique. We first add a source term  to $Z_>^0$:
\be
Z^0_>\to Z^0_> [\eta(\om_n)]  &=&  \int \mathcal{D} x_>~
e^{-\beta ~t~\sum_{|\om_n| \in S_b}\Big[ \frac{\gamma ~\om_n^2}{4} ~\x(\om_n)~ \x(-\om_n)~ +~\x(\om_n) \eta(-\om_{n})\Big]}\\
&=& \int \mathcal{D} x_>~
e^{-\frac{\beta~t}{2}~\sum_{|\om_n|\in S_b}\Big[\frac{\gamma~}{2} \om_n^2 
~\left(\x(\om_n) + \frac{2}{\gamma \om^2_n}~\eta(\om_n)~\right) 
~\left(\x(-\om_n) + \frac{2}{\gamma \om^2_n}~\eta(-\om_n)~\right)
+ \frac{2}{\gamma~\om_n^2}~\left(\eta(\om_{n})~\eta(\om_{-n}) \right)\Big]}
\ee 
Then, we  functionally differentiate twice with respect  to the source:
\be
\label{G0}
G^0_>(\omega_n, \omega_m) &=& 
\lim_{\eta\to 0}~\frac{1}{(\beta ~t)^2} ~\frac{\delta}{\delta
  \eta(-\om_n)} ~\frac{\delta}{\delta \eta(-\om_m)}  ~e^{\frac{\beta~t}{2} ~\sum_{|\om_n|\in S_b}~\eta(\omega_n) \eta(-\om_n) }=
~\frac{2}{\beta ~\gamma ~t~ \om_n^2}~\delta_{\om_m +\om_n,0} .
\ee
Note    that since the zero mode belongs to the slow modes part  of the kinetic  action, the kinetic operator for the fast modes is never singular and can be inverted without troubles. 

\section{Diffusion Monte Carlo algorithm}
\label{algo}
Our numerical study were performed using the following diffusion Monte Carlo algorithm: 
\begin{enumerate}
\item A ensemble of $N_w=18000$ initial configurations $\{ x_1(t=0), \ldots, x_{N_w}(t=0)\}$ was generated by sampling from  a narrow Gaussian distributions of width $\sigma=0.01$, centered at the origin $x=0$. Each of such positions represents the starting point of a walker.
\item A new set of $N_w$ configurations was obtained by evolving the initial points  for an elementary interval $\Delta~t$ , according to the Langevin dynamics in the Ito calculus:
\be
x_l(t+\Delta t) = x_l(t) - \frac{\Delta t}{\gamma}~\frac{d}{dx}~U(x_l(t))  + \Delta t \eta, \qquad l=1,\ldots, N_w.
\ee 
$ \Delta t \eta$ represents the usual Brownian diffusion term, which was performed by sampling from a Gaussian of width $\sigma^2=  \frac{2}{\beta \gamma} \Delta t$,  centered at the origin. 
\item For each walker, we generated a  random number $\xi\in[-0.5,0.5]$  and we made $N_c$ copies of the walker, where $N_c$  is the integer part of 
$ \mathcal{W}(x(t+\Delta t))+\xi$.
Hence, for $N_c=0$ the  walker was aborted, for  $N_c=1$ the walker was left unchanged, while for $N_c>1$ the walker gave raise to descendents, which then propagated independently from the progenitor. 
The integration time step $\Delta t$  was chosen in such a way that the relative fluctuations in the 
 population of walkers was only occasionally exceeding  $10\%$.
 \item The steps 2-3  were iterated for many integration time steps. 
 \item The quantity $\langle x(t) \rangle$ was obtained from the mean over the configurations of the walkers.  The statistical error was estimated from the variance.
\end{enumerate} 

\acknowledgments

We thank A.Szabo for reading the manuscript and making useful comments. Discussions with P.Armani, M.Sega, F.Pederiva, P.Verrocchio and G.Garberoglio were also useful. P.F. acknowledges financial support from C.N.R.S., during his permanence at the I.Ph.T. of C.E.A. and from I.N.F.N., under the AD31 scientific initiative. The work of O.C. was partly supported by the Italian MIUR-PRIN contract 20075ATT78.

\end{document}